\documentclass[10pt,letterpaper]{article}

\usepackage{cogsci}
\usepackage{pslatex}
\usepackage{apacite}
\usepackage{amsmath}
\usepackage{graphicx}

\title{Learning Criticality in an Embodied Boltzmann Machine}
 
\author{{\large \bf Miguel Aguilera (sci@maguilera.net)} \\
  Dept. of Computer Science, Univ. of Zaragoza, Zaragoza (Spain) \\
  Department of Psychology, University of the Balearic Islands, Palma (Spain)\\
  Aragon Institute of Engineering Research, Zaragoza (Spain)
\AND {\large \bf Manuel G.~Bedia (mgbedia@unizar.es)} \\
    Dept. of Computer Science, Univ. of Zaragoza, Zaragoza (Spain) \\
    Aragon Institute of Engineering Research, Zaragoza (Spain)}

\begin{document}

\maketitle

\begin{abstract}

Many biological and cognitive systems do not operate deep into one or other regime of activity. Instead, they exploit critical surfaces poised at transitions in their parameter space. 
The pervasiveness of criticality in natural systems suggests that there may be general principles inducing this behaviour.
However, there is a lack of conceptual models explaining how embodied agents propel themselves towards these critical points. In this paper, we present a learning model driving an embodied Boltzmann Machine towards critical behaviour by maximizing the heat capacity of the network. 
We test and corroborate the model implementing an embodied agent in the mountain car benchmark, controlled by a Boltzmann Machine that adjust its weights according to the model. We find that the neural controller reaches a point of criticality, which coincides with a transition point of the behaviour of the agent between two regimes of behaviour, maximizing the synergistic information between its sensors and the hidden and motor neurons.
Finally, we discuss the potential of our learning model to study the contribution of criticality to the behaviour of embodied living systems in scenarios not necessarily constrained by biological restrictions of the examples of criticality we find in nature.


\end{abstract}

\section{Introduction}
The application of notions of self-organization and complex systems to psychology and cognitive science has increased during the last decades. Recently, some contributions in this field have obtained sets of indicators of critical behaviour (long-range correlations, fractal scaling, etc.)  to characterize  different domains of cognitive activity \cite{chialvo_emergent_2010, van_orden_blue-collar_2012, dixon_multifractal_2012}.
Although comparing experimentally measurable quantities with the parameters of models of self-organized criticality allows to establish analogies between models and observed phenomena, the connection between the empirical indicators and mechanistic models is often thin \cite{wagenmakers_abstract_2012}.
Mechanistic theories, accounting for the behaviour of the system in terms of its parts and their interactions \cite{bechtel_discovering_2010} can provide useful perspectives on the study of complex notions as criticality that the majority of current empirical explanations do not have incorporated.
 
Interestingly, in the past few years, large sets of biological real data have allowed to characterize -using mechanistic models- how the behaviours of different biological systems (e.g. networks of neurons, antibody segments or flocks of birds) are poised near a critical point in their parameter space \cite{mora_are_2011}.
This is a great step towards the development of deeper theoretical principle behind the behaviour of biological and cognitive system.
However, beyond the importance of these models explaining the emergence of criticality in specific experimental data, we suggest that a complementary perspective could tackle the development of `conceptual models' in order to explain how organisms are driven towards critical behaviour in a more abstract level.
Conceptual models are defined in terms of a set of general mechanisms and generic processes but expressed in abstract frameworks working as `proofs of concept' \cite{barandiaran_animats_2009} that can be the support of future  experiments.

% In this paper we are interested in analyzing the role of the criticality from this `abstract perspective'. We believe that the study of the notion of `criticality
% at a theoretical level' can provide a better understanding of the reasons for which fractal measures are universally found in different psychological tasks and in the 
% cognitive processes involved.

% \section{Criticality from a statistical modelling perspective}

In this paper, we propose a conceptual model describing a mechanisms driving an embodied agent towards critical points of its parameter space. 
We make use of  concepts from statistical mechanics to base our model not on specific configurations of the parameters of the agent, but instead we exploit macroscopic variables to drive the system to transition points between qualitatively different regimes of behaviour. 
Driving synthetic agents to criticality may offer the opportunity to clarify what is its contribution of in different contexts. In the study of cognitive processes, we always find that criticality appears entangled with other features of adaptive behaviour (e.g. perception, prediction, learning) in agents dealing with complex environments. A mechanism poising agents in criticality in different scenarios may help understanding what are the contributions of criticality `by-itself' or how is criticality related to other phenomena.

In order to do it, we first introduce a Boltzmann Machine as the simplest statistical mechanics model showing correlations between elements of a network and derive a learning model driving the system towards critical points. The model will exploit the heat capacity of the system, as a macroscopic property that works as a proxy for criticality (when the heat capacity diverges the Boltzmann Machine is in a critical point).
After that, we test our learning model in an embodied agent controlling a Mountain Car (a classic reinforced learning testbed) finding that it is able to drive both the neural controller and the behaviour of the agent to a transition point in the parameter space between qualitatively different behavioural regimes.
Finally, we discuss the possible applications of our model to contribute to the development of deeper principles governing biological and cognitive systems.

% To facilitate the analysis of the next section, we introduce some of the core concepts of statistical modelling. The simplest stochastic model, from a computational framework, is called a Boltzmann network and works as a bridge between (i) a microdescription in terms of the possible values of its variables satisfying the constraints of the problem and (ii) the macroproperties of the system. In a Boltzmann network of N neurons and energy E, it is assumed that the most probable distribution of neurons in each energy level can be obtained by maximising an appropriate probability function which can be essentially understood as the entropy function.
% 
% FALTA: concluir (relacion con Ising) y cerrar intro.

\section{Driving a neural controller towards a critical point}

We propose a learning model self-organizing the parameters of a Boltzmann Machine, in order to drive the system towards states of criticality.
We take advantage of the fact that at critical points, derivatives of thermodynamic quantities as the entropy may diverge \cite{mora_are_2011}. An example of this is the heat capacity, whose divergence is a sufficient condition for criticality (though not a necessary one).
We define our network as a Boltzmann Machine \cite{ackley_learning_1985} following a maximum entropy distribution:
\begin{equation}
 P(s) = \frac{1}{Z}e^{-\beta E(s)}
\end{equation}
where the energy of each state is defined in terms of the bias and couplings of the state of each neuron.
\begin{equation}
 E(s)=-(\sum_i h_i s_i + \sum_{i,j} J_{ij}s_i s_j)
\end{equation}
The states $s_i$ can take values of $+1$ or $-1$ and the couplings $J_{ij}$ and bias $h_i$ can take continuous values.

In order to define a learning rule adjusting the values of $h_i$ and $J_{ij}$ we define a gradient climbing rule for maximizing the value of the heat capacity, with the intent of driving the system to critical points depicted by a singularity of the heat capacity.
Though, the heat capacity of the global state of the system depends on global variables of the system (e.g. the energy of the system) and thus we cannot define a gradient climbing rule based only in local information. Instead, we can define the heat capacity of the system from the path entropy of each neuron depicting transition between states, which is defined by the probability:
\begin{equation}
 P(s_i'|s) = \frac{e^{\beta_i s_i'H_i}}{2cosh(\beta H_i)},\qquad H_i=h_i+\sum_j J_{ji}s_j
\label{eq:Boltzmann-distribution}
\end{equation}
where $s$ is the state of the system at time $t$ and $s'$ at time $t+1$ and $\beta_i$ is now ascribed to the transitions of an individual neuron\footnote{Using individual values of $\beta_i$ allows to derive a learning rule that is only dependent on local variables}.
Path entropy is defined as the entropy of the transitions of the state of a neuron $i$,
\begin{equation}
\begin{split}
 S(s_i'|s) = -\sum_{s} P(s) \sum_{s_i'} log(P(s_i'|s))\cdot P(s_i'|s) = \\ = -\sum_{s} P(s) (\beta_i H_i tanh(\beta_i H_i) - log(2 cosh(\beta_i H_i))
 \end{split}
 \label{eq:entropy}
\end{equation}
From the path entropy we can define the heat capacity associated with the path entropy of neuron $i$ as
\begin{equation}
\begin{split}
 C_i = -\beta_i \frac{\partial S(s'_i|s)}{\partial\beta_i} = \sum_s  P(s) (\frac{H_i^2 \beta_i^2}{cosh(\beta_i H_i)^2} + \\ +
   \beta_i (s_i H_i - \langle s_i H_i \rangle)(\beta_i H_i tanh(\beta_i H_i) - log(2 cosh(\beta_i H_i)))
 \end{split}
 \label{eq:heat-capacity}
\end{equation}

In our model, the value of the thermodynamic beta defines the temperature of the system $\beta_i = \frac{1}{k_B T_i}$, where $k_B$ is the Boltzmann constant and $T_i$ the temperature associated with each neuron. Nevertheless, since the temperature here has no real-world meaning, $\beta_i$ just corresponds to a global rescaling of the parameters of the neuron by multiplying them by a constant value. Thus, we determine a working temperature defining $\beta_i=1$.

Considering $F_i = H_i tanh(H_i) - log(2 cosh(H_i)$, $G_i = \frac{H_i^2}{cosh(H_i)^2} + s_i H_i F_i$ and $K_i =\langle s_i H_i \rangle$ and knowing that $\frac{\partial P(s)}{\partial h_i} = (s_i- \langle s_i \rangle) P(s)$ and  $\frac{\partial P(s)}{\partial J_{ij}} = (s_i s_j - \langle s_i s_j \rangle) P(s)$ we derive the learning rules that climb the gradient of $C_i$ and drive the system towards critical points as:

\begin{equation}
\begin{split}
 \frac{\partial C_i}{\partial h_i} = \langle \frac{\partial G_i}{\partial h_i} \rangle + \langle s_i G_i \rangle - \langle s_i \rangle \langle G_i \rangle - \frac{\partial K_i}{\partial h_i} \langle F_i \rangle - \\ - K_i (\langle \frac{\partial F_i}{\partial h_i} \rangle + \langle s_i  F_i \rangle - \langle s_i \rangle \langle F_i \rangle)
 \\
 \frac{\partial C_i}{\partial J_{ji} }= \langle \frac{\partial G_i}{\partial J_{ji}} \rangle + \langle s_i s_j G_i \rangle - \langle s_i s_j \rangle \langle G_i \rangle - \frac{\partial K_i}{\partial J_{ji}} \langle F_i \rangle - \\ - K_i (\langle \frac{\partial F_i}{\partial J_{ji}} \rangle + \langle s_i s_j F_i \rangle - \langle s_i s_j \rangle \langle F_i \rangle)
 \end{split}
 \label{eq:learning}
\end{equation}
where
\begin{equation}
 \begin{split}
  \frac{\partial F_i}{\partial h_i} = \frac{H_i}{cosh(H_i)^2} , \\
  \frac{\partial F_i}{\partial J_{ji}} = \frac{H_i s_j}{cosh(H_i)^2} ,   \\
  \frac{\partial G_i}{\partial h_i} = \frac{2 H_i (1-H_i tanh(H_i))}{cosh(H_i)^2} + s_i F_i + s_i H_i \frac{\partial F_i}{\partial h_i} ,  \\
  \frac{\partial G_i}{\partial  J_{ji}} = \frac{2 H_i s_j (1-H_i tanh(H_i))}{cosh(H_i)^2} + s_i s_j F_i + s_i H_i \frac{\partial F_i}{\partial  J_{ji}} , \\
  \frac{\partial K_i}{\partial h_i} = \langle s_i \rangle + \langle s_i^2 H_i \rangle - \langle s_i \rangle K_i \\
  \frac{\partial K_i}{\partial  J_{ji}} = \langle s_i s_j \rangle + \langle s_i^2 s_j H_i \rangle - \langle s_i s_j \rangle K_i
 \end{split}
\end{equation}

In the following section, we use this learning rule  to drive the neural controller of an embodied agent towards a critical point. In order to do so, we need to take into account the environment during learning. If we consider two interconnected Boltzmann Machines, (one being the neural controller and other being the environment) Equation \ref{eq:learning} holds perfectly if we only apply it to the values of $i$ and $j$ corresponding to units of the neural controller.
In our case, we will not use a Bolzmann Machine as an environment but instead we will use a classic example from reinforced learning. Therefore, our learning rule will be valid as long as the statistics of the environment can be approximated by a Boltzmann Machine with an arbitrary number of units. Luckily, Boltzmann Machines are universal approximators \cite{montufar_universal_2014}. Nevertheless, if the updating of units does not follow the rules of a Boltzmann Machine\footnote{For example, in our embodied model sensors values are clamped from values in the environment, thus they influence hidden and motor units but they are not influenced by them. In a Boltzmann Machine, this can provoke that the distribution of states no longer follows the Boltzmann distribution depicted by Equation \ref{eq:Boltzmann-distribution}}, as we will see later, the approximation can be flawed in some cases.

\section{Embodied model: Mountain Car}

\begin{figure}[ht]
\begin{center}
 \includegraphics[width=6.5cm]{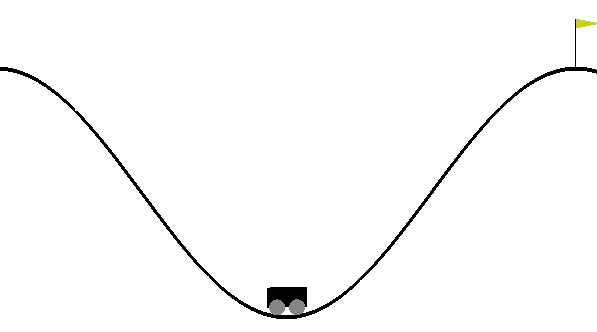}
\end{center}
\caption{Mountain Car environment. An  under-powered car that must drive up a steep hill by balancing itself to gain momentum.} 
\label{fig:MountainCar}
\end{figure}

In order to evaluate the behaviour of the proposed learning model, we test it in the Mountain Car environment \cite{moore_efficient_1990}. This environment is a classical test bed in reinforced learning depicting an under-powered car that must drive up a steep hill (Figure \ref{fig:MountainCar}). Since gravity is stronger than the car's engine, the vehicle must learn to leverage potential energy by driving to the opposite hill before the car is able to make it to the goal at the top of the rightmost hill. We simulate the environment using the OpenAI Gym toolkit \cite{brockman_openai_2016}.
In this environment, the horizontal position $x$ of the car is limited to an interval of $[-1.5\pi,0.5\pi]$, and the vertical position of the car is defined as $y=sin(3x)$. The velocity in the horizontal axis is updated each time step as $v(t+1)=v(t) + 0.001 a - 0.0025 cos(3x)$, where $a$ is the action of the motor which can be either ${-1,0,1}$. 
 
So as to make it difficult for random agents to solve the task, the maximum velocity of the car was limited to $[-0.045,0.045]$\footnote{Typically the Mountain Car environment restricts the velocity of the car to the interval $[-0.07,0.07]$}.
With this velocity limitation, only around a fraction of $6.30\%$ of agents with random parameters (sampled from a uniform distribution in the range $[-1,1]$) are able to reach the top of the mountain in a trial of $1000$ simulation steps starting from a starting random position (uniformly distributed in the region of the valley $[0.4,0.6]$).

We define the neural controller of the car as a Boltzmann Machine containing $6$ sensors and $6$ neurons. We feed the sensors with the horizontal and vertical acceleration of the car, each discretized to arrays of three bits. Each sensor unit is assigned a value of $1$ if its corresponding bit is active and $-1$ otherwise. Two of the car neurons are connected to the motors, defining $a=1$ if both neurons are active, $a=-1$ if both neurons are inactive, and $a=0$ otherwise.
We apply the learning rule from Equation \ref{eq:learning} to $10$ different agents. In order to avoid overfitting, we add an L2 regularization term updating the parameters of the system according to the rule:
\begin{equation}
 \begin{split}
  h_{i} \leftarrow h_{i} + \mu  \frac{\partial C_i}{\partial h_{i} } - \lambda  h_{i} \\
  J_{ji} \leftarrow J_{ji} + \mu  \frac{\partial C_i}{\partial J_{ji} } - \lambda  J_{ji}
 \end{split}
\end{equation}
where $\mu=0.02$, $\lambda=0.002$ and $\frac{\partial C_i}{\partial h_{i} }$ and $\frac{\partial C_i}{\partial J_{ji} } $ are the result of Equation \ref{eq:learning}.
Agents are initialized in the starting random position of the environment. Hidden and motor neurons are randomized and the initial parameters $h$ and $J$ are sampled from a uniform random interval $[-0.01,0.01]$. The agents are simulated for $1000$ trials of $5000$ steps, applying Equation \ref{eq:learning} at the end of the trial computing the values of $\frac{\partial C_i}{\partial h_{i} }$ and$\frac{\partial C_i}{\partial J_{ji} } $ over that trial. Note that agents are not reseted at the end of the trial.

\section{Results}

In this section, we analyze the behaviour of the neural controllers and the behavioural patterns of the agents respect the possibilities of their parameter space. Although figures correspond to one particular agent (one of the ones reaching the top of the mountain), most results are general to all 10 agents, except when it is indicated otherwise. In order to compare the agents with other behavioural possibilities, we explore the parameter space by changing the parameter $\beta$ of the agents. Modifying the value of $\beta$ is equivalent to a global rescaling of the parameters of the agent transforming $h_{ji} \leftarrow  \beta \cdot h_{ji}$ and $J_{ji} \leftarrow \beta \cdot J_{ji}$, thus exploring the parameter space along one specific direction. For $21$ values of $\beta$ logarithmically distributed in the interval $[10^{-1},10^1]$ we simulate the 10 agents for a trial of $10^6$ simulation steps, after starting the agents from the random starting position (i.e. $x$ in an interval $[0.4,0.6]$) and a initial 
run of $10^4$ simulation steps. We will use the results of those simulations for all the results in this section.

\subsection{Signatures of criticality in the neural controller}
Firstly,  we test whether the trained agents show signatures of critical behaviour. 
Counting the occurrence of each possible state of the $12$ neurons of the agents (including sensor, hidden and motor neurons) we can compute the probability distribution of the Boltzmann Machine $P(s)$.

\begin{figure}[ht]
\begin{center}
 \begin{tabular}{ll}
  \textbf{A} & \textbf{B} \\
 \includegraphics[width=4cm]{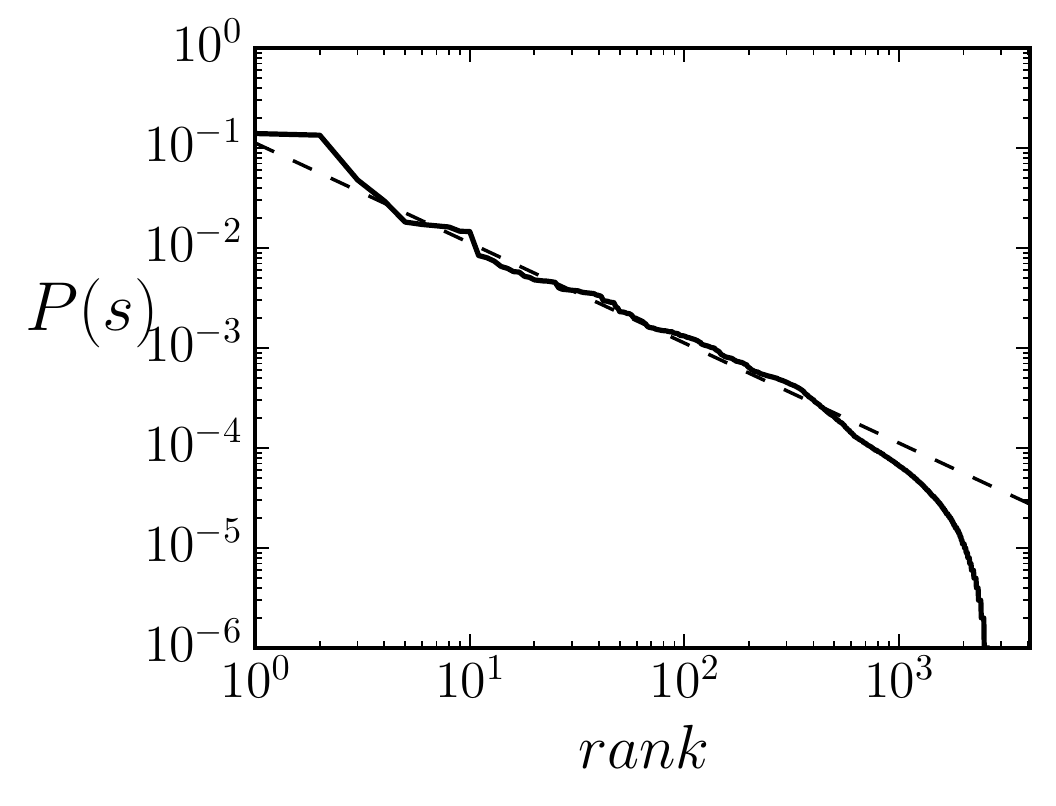} &
 \includegraphics[width=4cm]{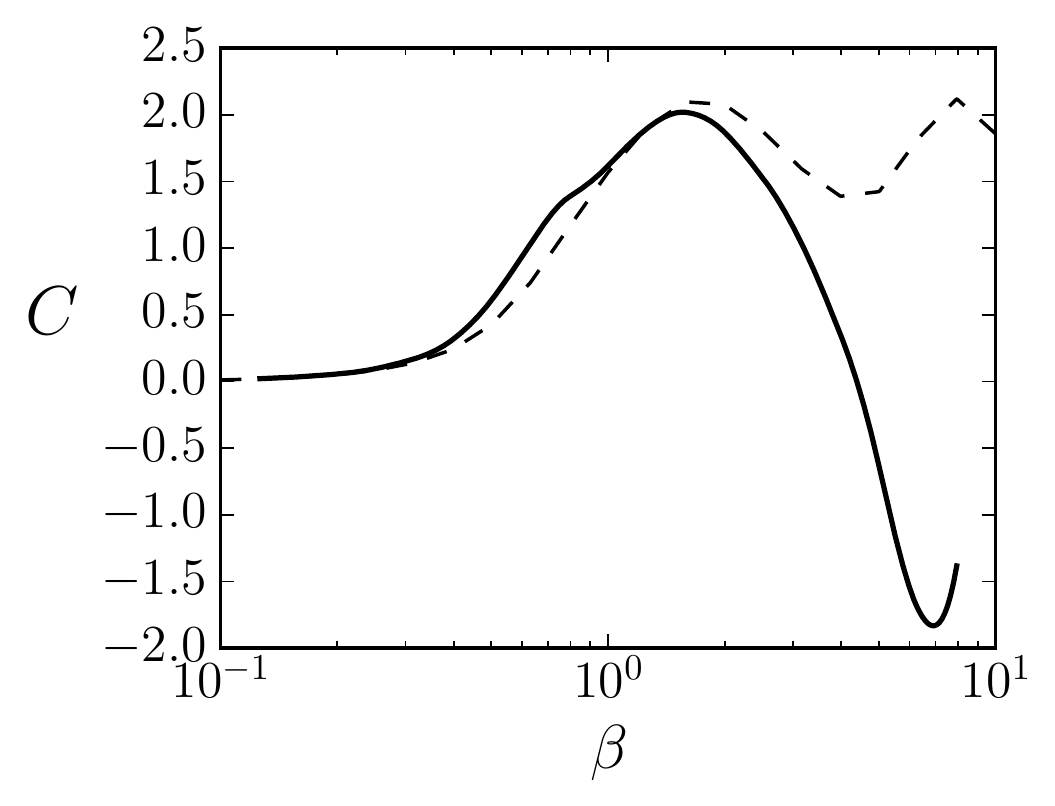}
 \end{tabular}
\end{center}
\caption{\textbf{Signatures of criticality}.  \textbf{(A)} Ranked probability distribution function of the inferred Ising models for the different frequency bands (solid line) versus a distribution following Zipf's law, (i.e. $P(s) = 1/rank$, dashed line). We observe a good agreement between the model and Zipf's law, suggesting critical scaling. \textbf{(B)} Heat capacity versus $\beta$ computed using Equation \ref{eq:entropy} for calculating the entropy and derivating a cubic interpolation of the entropy function respect to $\beta$ (solid line) and estimation of the heat capacity using the approximation used in Equation \ref{eq:heat-capacity} (dashed line). A peak in heat capacity is observed near $\beta=1$, suggesting that the system is near a critical point. For values of $\beta$ below the critical point we observe that the heat capacity and its approximation coincide, indicating that the approximation is valid for that range.}
\label{fig:criticality}
\end{figure}

We observe that all agents approximately follow a Zipf's law at $\beta=1$ (Figure \ref{fig:criticality}.A) for almost three decades, which is a good agreement for the limited size of the system (note that the possible states of the system are limited to $2^{12}$ states. All trained agents show a similar distribution close to Zipf's law.

Secondly, as another indicator of critical points is the divergence of the heat capacity of the system, we estimate the heat capacity of hidden and motor neurons \footnote{ We can do it using Equation \ref{eq:entropy} or,   alternatively, using Equation \ref{eq:heat-capacity}, which involves the approximations made for designing the learning algorithm that models heat capacity `as seen' by the neural controller. }. 
The result that we observe (Figure \ref{fig:criticality}.B) is that the heat capacity peaks around the operating temperature (at a value slightly larger than $\beta=1$) that, together with the Zipf's distribution, it suggests that the system is operating in a regime of criticality.

Finally, if we  compare the real heat capacity and the heat capacity as seen by the learning algorithm, we can infer that the approximation of the environment as a Boltzmann Machine works well when the parameters of the agent are not too large (increasing $\beta$ is equivalent to rescale all parameters of the system similarly, and therefore, regularization terms might be necessary for the learning algorithm to work correctly at least when environments are deterministic, as in this case).

\begin{figure}[]
\begin{center}
 \begin{tabular}{lll}
  \textbf{A} & \textbf{B} & \textbf{C} \\
 \includegraphics[width=2.7cm]{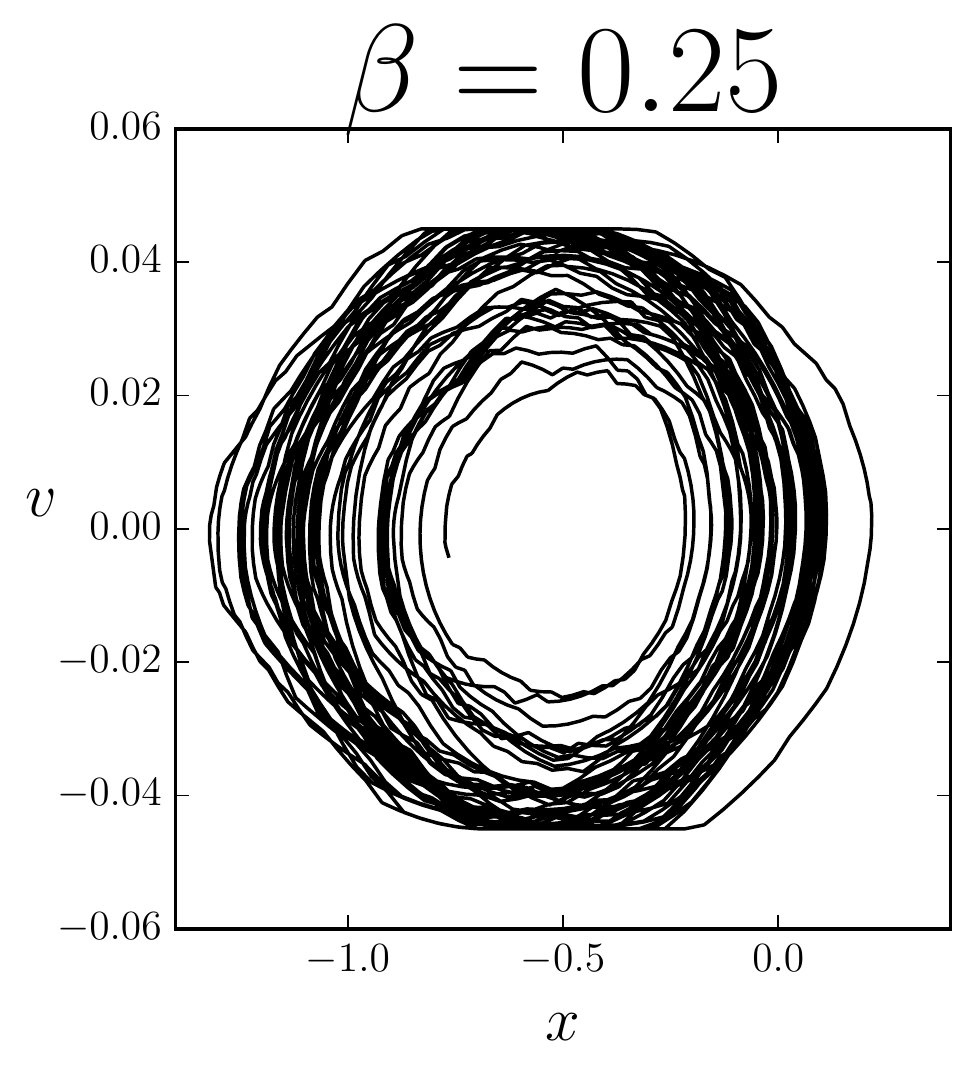} &
 \includegraphics[width=2.7cm]{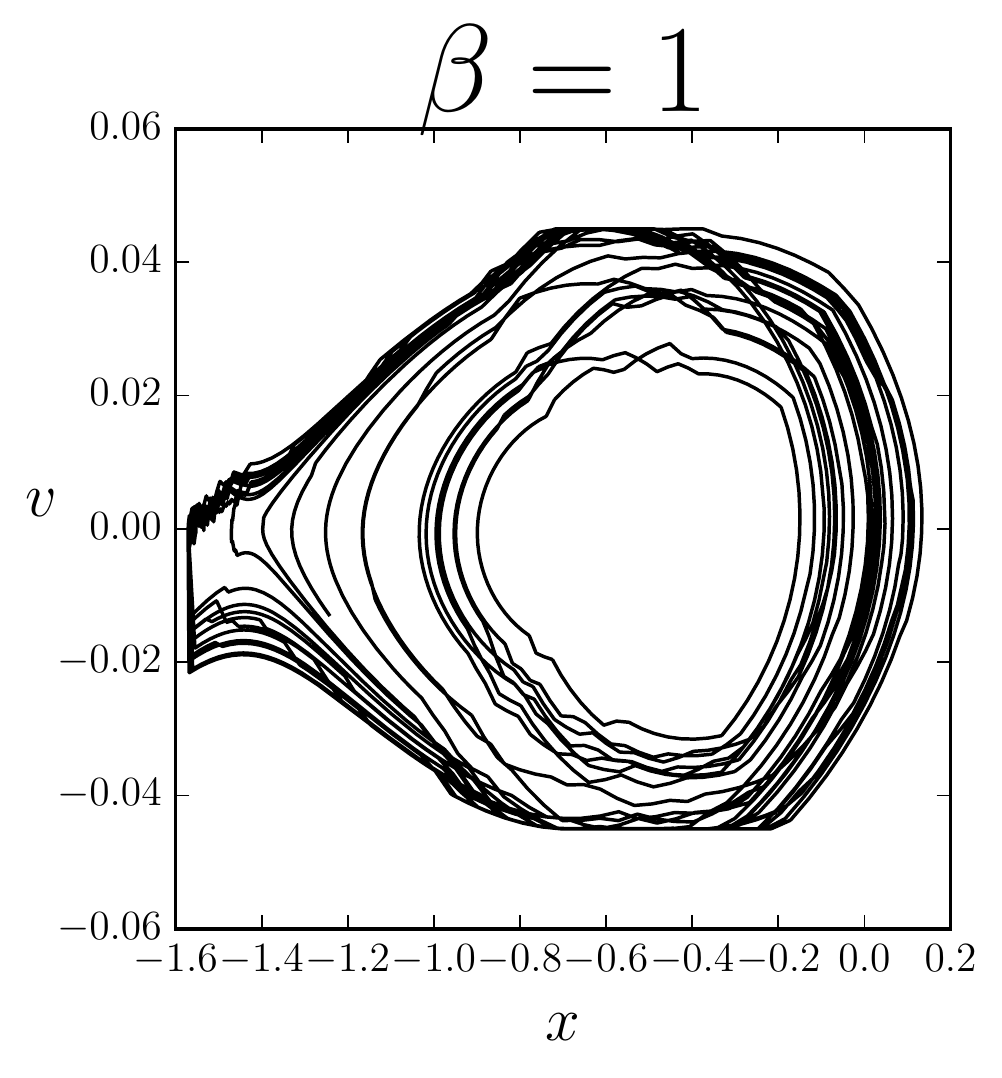} &
 \includegraphics[width=2.7cm]{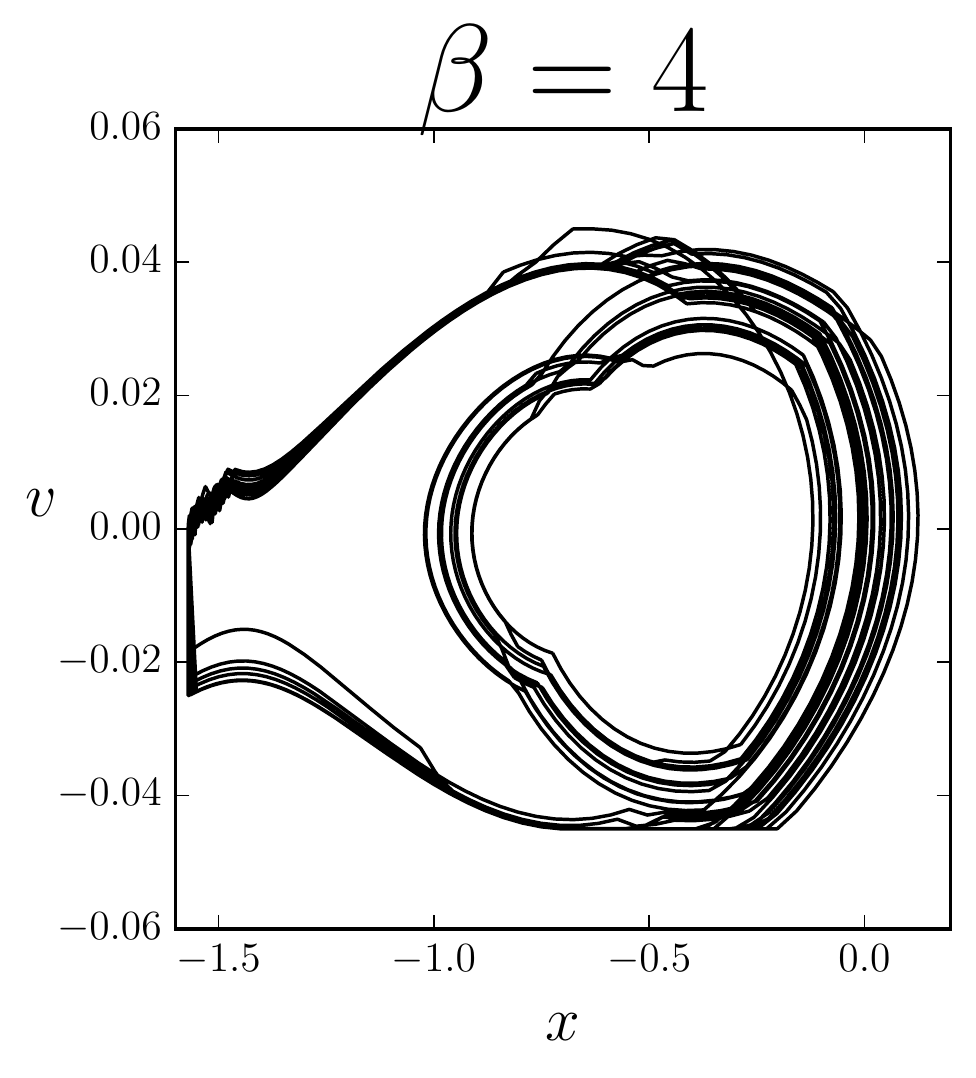} \\
 \includegraphics[width=2.7cm]{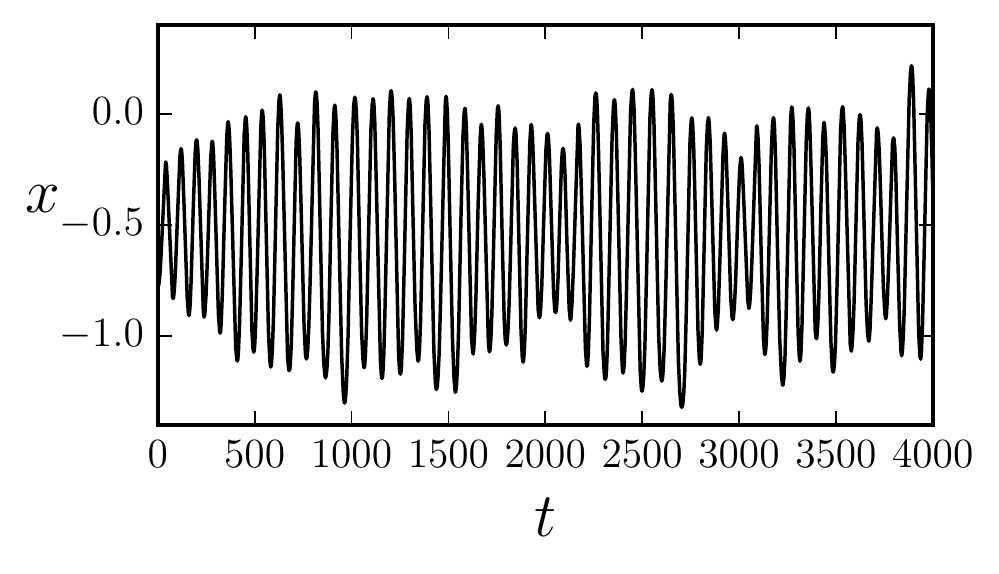} &
 \includegraphics[width=2.7cm]{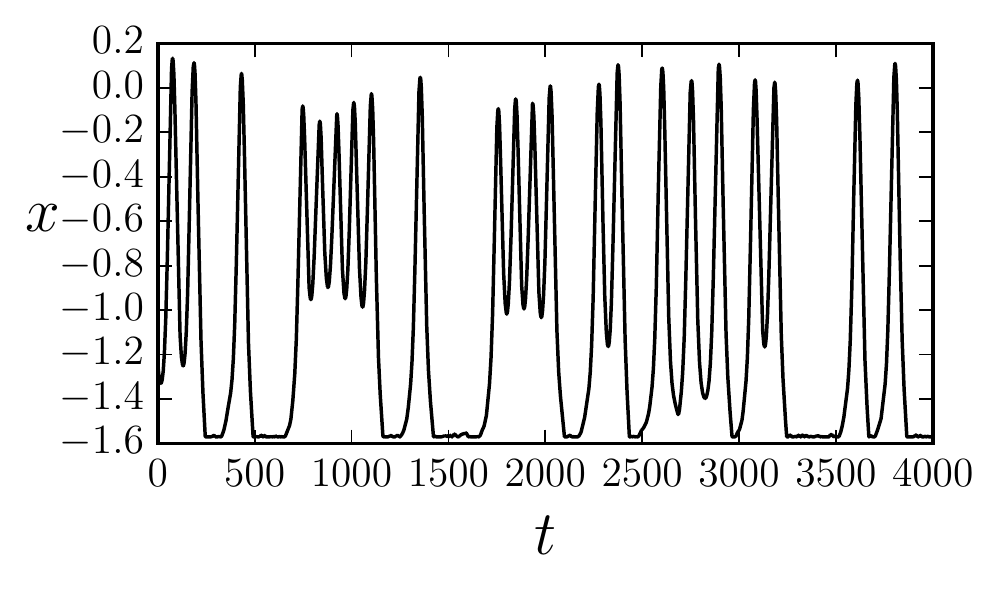} &
 \includegraphics[width=2.7cm]{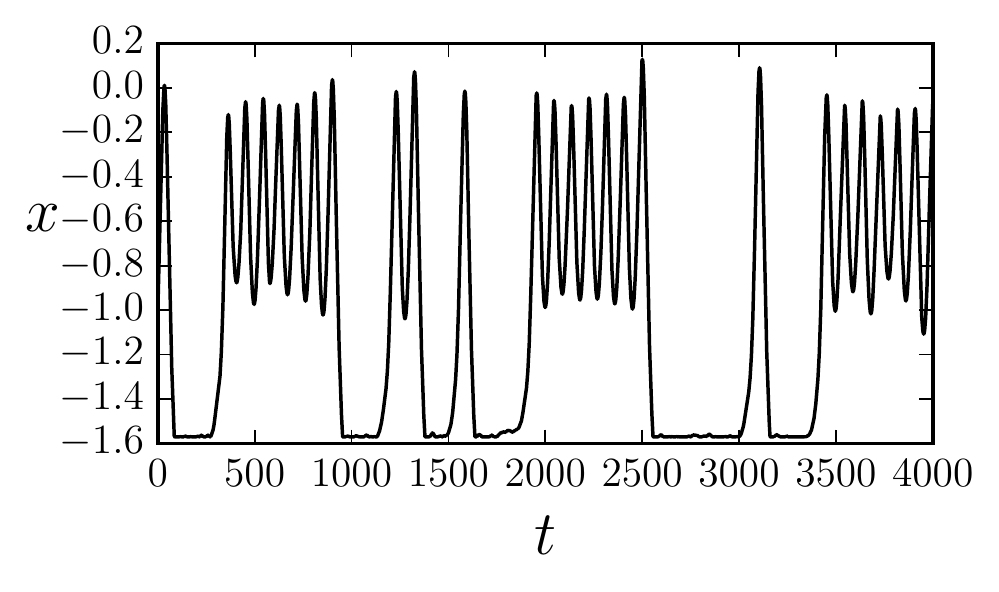}
 \end{tabular}
\end{center}
 \textbf{D}
\begin{center}
 \includegraphics[width=9cm]{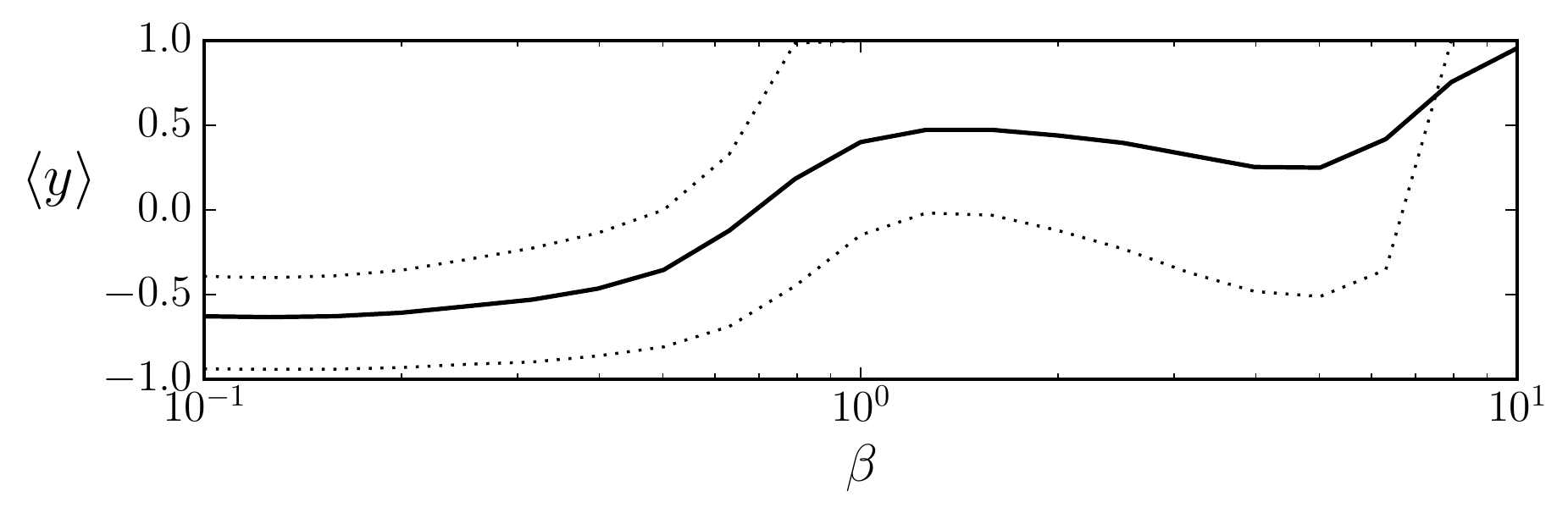}
\end{center}
\caption{Transition in behavioural regime of the agent. We show the behaviour of an agent for an interval of $4000$ steps with values of $\beta$ of $0.25$ (\textbf{A}), $1$ (\textbf{B}) and $4$ (\textbf{C}), depicting the trajectories of the car in its phase space ($x$ versus $v$, top) and the evolution of the values of $x$ (bottom). We observe that $\beta=1$ is a transition point between two modes of behaviour. (\textbf{D}) Average vertical position of the car $\langle y \rangle$ (solid line) and its upper and lower quartiles (dotted lines). We observe a transition near $\beta=1$ where the agent reaches the top of the mountain. Similar transitions are identified in $6$ of the $10$ simulated agents.}
\label{fig:transitions}

\end{figure}
\begin{figure}[]
\begin{center}
 \includegraphics[width=7.5cm]{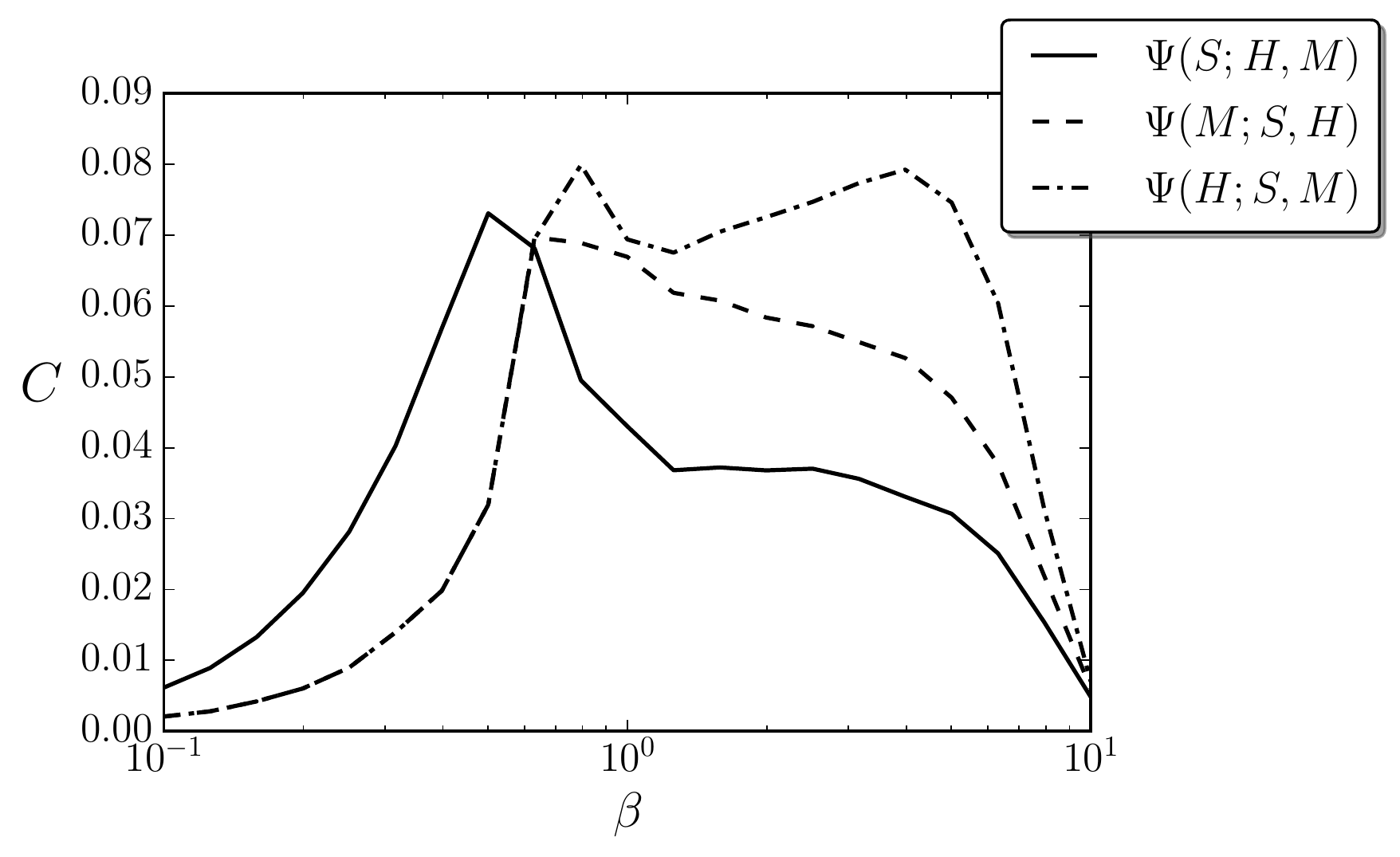}
\end{center}
\caption{Synergistic information. Synergy values $\Psi(S;M,H)$ (solid line), $\Psi(M;S,H)$ (dashed line) and $\Psi(H;S,M)$ (dash-dot line) computed using partial information decomposition. We find a transition from high synergy between motor and hidden neurons about sensory information to high synergy between sensor neurons and motor and hidden neurons respectively near $\beta=1$.} 
\label{fig:synergy}
\end{figure}

\subsection{Behavioural transitions in the parameter space}
What does it imply for the agent to poise its neural controller at a critical point?

It should be remarked here that our agents are given no explicit goal but they only tend to behavioural patterns maximizing the heat capacity of its neurons independently of this behaviour reaches the top of the mountain or not (in fact, only $4$ of the $10$ trained agents are able to climb to the top of the mountain). 
Related to this, we will start exploring the effects of transiting the critical point observing the different behavioural modes of the agent in the parameter space. 
The behaviour of the car can be described just by the position $x$ and speed $v$ at different moments of time.

In Figure \ref{fig:transitions}.A-C we can observe the behaviour of the car for $\beta=0.25,1,4$ respectively, for an interval of $4000$ simulation steps. If we compute the average value of $y$ at the trial for each value of $\beta$ Figure \ref{fig:transitions}.D, we observe that slightly below the operating temperature there is a transition from agents that are not able to reach to the top of the mountains to those that are able to do so. 
More in detail, in the $4$ agents that are able to reach the top of the mountain the results are similar while $2$ more agents that are not able to reach the top display a similar transition in the average value of height $y$ and the average absolute velocity $v$ of the car. The remaining $4$ agents do not show a transition in average values of basic behavioural variables although it does not preclude the possibility of another type of less evident behavioural transition.

What is it changed in this behavioural transition? We are interested in knowing of these behavioural regimes are qualitatively different and  will explore this using information theory to characterize how different variables of the agent interact at different points of the parameter space. Specifically, we are interested in the relation between sensor, hidden and motor neurons, which determines the behaviour of the agent in its environment. 

But, is the agent merely reactive to sensory inputs or is there a more complex interplay between sensor, hidden and motor units? In order to answer this, we characterize the interaction between variables using partial information decomposition \cite{timme_synergy_2014} to compute synergies between variables, defined as:
\begin{equation}
 \begin{split}
 \Psi(Y;X_1,X_2)  = I(Y;X_1,X_2) - I(Y;X_1) - \\ -  I(Y;X_2) + I_{min}(Y;X_1,X_2)
 \end{split}
\end{equation}
where $Y$, $X_1$ and $X_2$ are random discrete variables, $I(Y;X)$ is the mutual information between two variables and $I_{min}(Y;X_1,X_2)$, defined as in \cite{williams_generalized_2011}, is the redundant information that $X_1$ and $X_2$ share about $Y$. The resulting synergy $\Psi(Y;X_1,X_2)$ is able to capture information of $Y$ that is not available from either $X_1$ and $X_2$  alone but from their interaction (the classical example is the relation between the output and inputs of an XOR gate).

Defining $S$, $H$ and $M$ as the joint distribution of sensor, hidden and motor neurons respectively, we can analyze the synergistic information between the distribution of the three groups of variables. The objective is to capture how much information emerges from the interaction between variables instead of being contained in the variables alone.

As we observe in Figure \ref{fig:synergy}, the synergy $\Psi(S;H,M)$ between motor and hidden neurons about sensor information peaks at values of $\beta$ lower than 1, while the synergies of hidden and motor neurons with sensor neurons, $\Psi(M;S,H)$ $\Psi(H;S,M),$ increase at larger values of $\beta$, depicting a transition point at a value of $\beta$ slightly lower than $1$. Since the environment of the agent is completely deterministic, it seems adequate that larger values of $\beta$ (i.e. less random behaviour) are more effective to transmit information from sensors to other neurons, while maximum interaction between hidden and motor neurons takes place at a place with a lower $\beta$.

\section{Discussion}

Recapitulating, we have proposed a learning model driving an embodied agent close to critical points in the parameter space, poising both the neural controller and the behavioural patterns of the agent near a transition point between qualitatively different regimes of operation.
In the case of the neural controller, we have found that the Boltzmann Machine has a peak in its peak capacity in a point slightly above $\beta=1$. 
However, if we analyze the synergistic interaction between sensor, hidden and motor units of the system we find a transition at a point slightly under $\beta=1$, which also coincides with a point of transition between behavioural regimens in $6$ of the $10$ agents. 
These results could suggest that the system might be finding a compromise between the critical point of its neural controller and zones of transition in the behavioural regimes of the agent (the former happening at higher temperatures and the latter at lower temperatures).

At this point, we could harken back to our original questions. Why do biological systems behave near to criticality? What are the benefits for a biological system to move to this special type of points? And more importantly, how can our learning model help answer those questions?
% What does criticality provide for natural systems?

Reviewing related literature, one finds that interpretations about criticality are too speculative in general. For example, \citeA{beggs_criticality_2008} hypothesizes that neural systems operating at a critical point can optimize information processing and its computational power. 
\citeA{mora_are_2011} discuss the experimental evidence of criticality in a wide variety of systems and propose that criticality could provide better defense mechanisms against predators (in animals), gain selectivity in response to stimuli (in auditory systems) or improved mechanism to anticipate attacks (in immunological systems). 
Nevertheless, the reasoning that give support to this hypothesis is based more on generic suggestions than on scientifically rigorous statements. 
% For example, authors conjecture that optimization of information processing could be related to the size of neuronal avalanches and the range of stable patterns, but without clear results that support these predictions. 
% The simple fact that those hypotheses can offer plausible descriptions cannot be the reason for automatically accepting them as explanations.
More detailed analyses are needed to accept speculations,and our opinion is that a conceptual model of embodied criticality in natural systems can be useful to capture how transition points in the parameter space of behavioral regimes can be found and exploited to obtain functional advantages as the ones mentioned above. For that purpose, rather than specific biological instances of critical phenomena, we have used an abstract framework of how embodied agents can be driven to critical points.

Furthermore, we believe that using conceptual models as the one presented here could us to test more intriguing hypothesis. 
For example, our general mechanism driving an embodied neural controller to criticality has the potential of capturing what is the contribution of criticality `by itself' to the behaviour of an adaptive agents in different scenarios, as well as the relation between criticality and other biological and cognitive phenomena.

On the other hand, criticality generally appears entangled other capabilities developed by biological systems, and typically interpretations about the advantages of criticality always refer to tangible benefits for the system (e.g. in an evolutive level, as the source of a new range of capabilities or better mechanism for surviving in open environments, etc.) and it is difficult to distinguish if criticality is the cause or the consequence of such effects.
Notwithstanding, as we have presented here, our model does not address any particular task. 
Instead, the model finds ways to drive the system to critical points, allowing us to explore what are the effects of poising a system to criticality under some embodied constraints, disentangling the effects of criticality to other factors embedded in real life organisms.
As well, this can be connected to the analysis particular features in animal behaviour that are interpreted without assuming a necessary pragmatical perspective of analysis. For example,  `play' in humans and other species does not aim to solve a specific problem, but instead it can be simply understood as a `rule-breaker' activity, breaking the constrains of stable regimes of behaviour, even if it is not directly required from the environment \cite{di_paolo_horizons_2010}.

Moreover, the presented model could be implemented in more complex embodied setups, for example involving specific tasks of adaptive behaviour adding environmental constraints (e.g. exploration, decision-making, categorical perception) or biological requirements (e.g. an internal metabolism or other biological drives as hunger or thirst) and observe how the compliance of these biological and cognitive requirements interplays with the drive towards critical points in the neural controller of the agent. Thus, we could explore in this way how criticality can contribute to capabilities observed by natural organisms.

The study of criticality in living systems has traditionally rested on too speculative grounds. Today, the increasing amount of high quality data together with the possibilities of statistical mechanics models promises exciting routes towards a rigorous exploration of the governing principles of biological organisms, linking experimental evidence and data-driven models with conceptual models exploring general mechanisms offering general explanations of the mechanisms driving the behaviour of these complex systems.
% In \cite{di_paolo_horizons_2010}, it is proposed an 
% explanation of play (in human children and other species) from a different perspective. Play do not solve any problem. Play is, in Di Paolo's words, `precisely not a 
% problem requiring a solution'. Authors as \citeA{donaldson_human_1993} formulates the benefits of play in a useful way in relation  to the improvement of socialization capabilities. 
% But the role of play also admits a non-pragmatic explanation. Play can be simply understood as a `rule-breaker' activity, the breaking of the constraints of stable and 
% self-equilibrating regime of behaviours, where it is not directly required from the environment. Play can be viewed as a way of making transitions between stable regimens generating norms without necessity. FALTA no instrumental

\section{Acknowledgments}

Research was supported in part by the Spanish National Programme for Fostering Excellence in Scientific and Technical Research project PSI2014-62092-EXP and for by the project TIN2011-24660 funded by the Spanish 
Ministry of Economy and Competitiveness.% Place acknowledgments (including funding information) in a section at

\bibliographystyle{apacite}

\setlength{\bibleftmargin}{.125in}
\setlength{\bibindent}{-\bibleftmargin}

\bibliography{references}

\begin{thebibliography}{}

\bibitem [\protect \citeauthoryear {%
Ackley%
, Hinton%
\BCBL {}\ \BBA {} Sejnowski%
}{%
Ackley%
\ \protect \BOthers {.}}{%
{\protect \APACyear {1985}}%
}]{%
ackley_learning_1985}
\APACinsertmetastar {%
ackley_learning_1985}%
\begin{APACrefauthors}%
Ackley, D\BPBI H.%
, Hinton, G\BPBI E.%
\BCBL {}\ \BBA {} Sejnowski, T\BPBI J.%
\end{APACrefauthors}%
\unskip\
\newblock
\APACrefYearMonthDay{1985}{}{}.
\newblock
{\BBOQ}\APACrefatitle {A learning algorithm for {Boltzmann} machines} {A
  learning algorithm for {Boltzmann} machines}.{\BBCQ}
\newblock
\APACjournalVolNumPages{Cognitive science}{9}{1}{147--169}.
\PrintBackRefs{\CurrentBib}

\bibitem [\protect \citeauthoryear {%
Barandiaran%
\ \BBA {} Chemero%
}{%
Barandiaran%
\ \BBA {} Chemero%
}{%
{\protect \APACyear {2009}}%
}]{%
barandiaran_animats_2009}
\APACinsertmetastar {%
barandiaran_animats_2009}%
\begin{APACrefauthors}%
Barandiaran, X\BPBI E.%
\BCBT {}\ \BBA {} Chemero, A.%
\end{APACrefauthors}%
\unskip\
\newblock
\APACrefYearMonthDay{2009}{{\APACmonth{07}}}{}.
\newblock
{\BBOQ}\APACrefatitle {Animats in the {Modeling} {Ecosystem}} {Animats in the
  {Modeling} {Ecosystem}}.{\BBCQ}
\newblock
\APACjournalVolNumPages{Adaptive Behavior}{17}{4}{287--292}.
\newblock
\begin{APACrefDOI} \doi{10.1177/1059712309340847} \end{APACrefDOI}
\PrintBackRefs{\CurrentBib}

\bibitem [\protect \citeauthoryear {%
Bechtel%
\ \BBA {} Richardson%
}{%
Bechtel%
\ \BBA {} Richardson%
}{%
{\protect \APACyear {2010}}%
}]{%
bechtel_discovering_2010}
\APACinsertmetastar {%
bechtel_discovering_2010}%
\begin{APACrefauthors}%
Bechtel, W.%
\BCBT {}\ \BBA {} Richardson, R\BPBI C.%
\end{APACrefauthors}%
\unskip\
\newblock
\APACrefYear{2010}.
\newblock
\APACrefbtitle {Discovering complexity: {Decomposition} and localization as
  strategies in scientific research} {Discovering complexity: {Decomposition}
  and localization as strategies in scientific research}.
\newblock
\APACaddressPublisher{}{MIT Press}.
\PrintBackRefs{\CurrentBib}

\bibitem [\protect \citeauthoryear {%
Beggs%
}{%
Beggs%
}{%
{\protect \APACyear {2008}}%
}]{%
beggs_criticality_2008}
\APACinsertmetastar {%
beggs_criticality_2008}%
\begin{APACrefauthors}%
Beggs, J\BPBI M.%
\end{APACrefauthors}%
\unskip\
\newblock
\APACrefYearMonthDay{2008}{}{}.
\newblock
{\BBOQ}\APACrefatitle {The criticality hypothesis: how local cortical networks
  might optimize information processing} {The criticality hypothesis: how local
  cortical networks might optimize information processing}.{\BBCQ}
\newblock
\APACjournalVolNumPages{Philosophical Transactions of the Royal Society of
  London A: Mathematical, Physical and Engineering
  Sciences}{366}{1864}{329--343}.
\PrintBackRefs{\CurrentBib}

\bibitem [\protect \citeauthoryear {%
Brockman%
\ \protect \BOthers {.}}{%
Brockman%
\ \protect \BOthers {.}}{%
{\protect \APACyear {2016}}%
}]{%
brockman_openai_2016}
\APACinsertmetastar {%
brockman_openai_2016}%
\begin{APACrefauthors}%
Brockman, G.%
, Cheung, V.%
, Pettersson, L.%
, Schneider, J.%
, Schulman, J.%
, Tang, J.%
\BCBL {}\ \BBA {} Zaremba, W.%
\end{APACrefauthors}%
\unskip\
\newblock
\APACrefYearMonthDay{2016}{}{}.
\newblock
{\BBOQ}\APACrefatitle {{OpenAI} gym} {{OpenAI} gym}.{\BBCQ}
\newblock
\APACjournalVolNumPages{arXiv preprint arXiv:1606.01540}{}{}{}.
\PrintBackRefs{\CurrentBib}

\bibitem [\protect \citeauthoryear {%
Chialvo%
}{%
Chialvo%
}{%
{\protect \APACyear {2010}}%
}]{%
chialvo_emergent_2010}
\APACinsertmetastar {%
chialvo_emergent_2010}%
\begin{APACrefauthors}%
Chialvo, D\BPBI R.%
\end{APACrefauthors}%
\unskip\
\newblock
\APACrefYearMonthDay{2010}{}{}.
\newblock
{\BBOQ}\APACrefatitle {Emergent complex neural dynamics} {Emergent complex
  neural dynamics}.{\BBCQ}
\newblock
\APACjournalVolNumPages{Nature Physics}{6}{10}{744--750}.
\newblock
\begin{APACrefDOI} \doi{10.1038/nphys1803} \end{APACrefDOI}
\PrintBackRefs{\CurrentBib}

\bibitem [\protect \citeauthoryear {%
Di~Paolo%
, Rohde%
\BCBL {}\ \BBA {} De~Jaegher%
}{%
Di~Paolo%
\ \protect \BOthers {.}}{%
{\protect \APACyear {2010}}%
}]{%
di_paolo_horizons_2010}
\APACinsertmetastar {%
di_paolo_horizons_2010}%
\begin{APACrefauthors}%
Di~Paolo, E\BPBI A.%
, Rohde, M.%
\BCBL {}\ \BBA {} De~Jaegher, H.%
\end{APACrefauthors}%
\unskip\
\newblock
\APACrefYearMonthDay{2010}{}{}.
\newblock
{\BBOQ}\APACrefatitle {Horizons for the enactive mind: {Values}, social
  interaction, and play} {Horizons for the enactive mind: {Values}, social
  interaction, and play}.{\BBCQ}
\newblock
\APACjournalVolNumPages{Enaction: Toward a new paradigm for cognitive
  science}{}{}{33--87}.
\PrintBackRefs{\CurrentBib}

\bibitem [\protect \citeauthoryear {%
Dixon%
, Holden%
, Mirman%
\BCBL {}\ \BBA {} Stephen%
}{%
Dixon%
\ \protect \BOthers {.}}{%
{\protect \APACyear {2012}}%
}]{%
dixon_multifractal_2012}
\APACinsertmetastar {%
dixon_multifractal_2012}%
\begin{APACrefauthors}%
Dixon, J\BPBI A.%
, Holden, J\BPBI G.%
, Mirman, D.%
\BCBL {}\ \BBA {} Stephen, D\BPBI G.%
\end{APACrefauthors}%
\unskip\
\newblock
\APACrefYearMonthDay{2012}{}{}.
\newblock
{\BBOQ}\APACrefatitle {Multifractal {Dynamics} in the {Emergence} of
  {Cognitive} {Structure}} {Multifractal {Dynamics} in the {Emergence} of
  {Cognitive} {Structure}}.{\BBCQ}
\newblock
\APACjournalVolNumPages{Topics in Cognitive Science}{4}{1}{51--62}.
\newblock
\begin{APACrefDOI} \doi{10.1111/j.1756-8765.2011.01162.x} \end{APACrefDOI}
\PrintBackRefs{\CurrentBib}

\bibitem [\protect \citeauthoryear {%
Mont\'ufar%
}{%
Mont\'ufar%
}{%
{\protect \APACyear {2014}}%
}]{%
montufar_universal_2014}
\APACinsertmetastar {%
montufar_universal_2014}%
\begin{APACrefauthors}%
Mont\'ufar, G\BPBI F.%
\end{APACrefauthors}%
\unskip\
\newblock
\APACrefYearMonthDay{2014}{}{}.
\newblock
{\BBOQ}\APACrefatitle {Universal {Approximation} {Depth} and {Errors} of
  {Narrow} {Belief} {Networks} with {Discrete} {Units}} {Universal
  {Approximation} {Depth} and {Errors} of {Narrow} {Belief} {Networks} with
  {Discrete} {Units}}.{\BBCQ}
\newblock
\APACjournalVolNumPages{Neural Computation}{26}{7}{1386--1407}.
\PrintBackRefs{\CurrentBib}

\bibitem [\protect \citeauthoryear {%
Moore%
}{%
Moore%
}{%
{\protect \APACyear {1990}}%
}]{%
moore_efficient_1990}
\APACinsertmetastar {%
moore_efficient_1990}%
\begin{APACrefauthors}%
Moore, A\BPBI W.%
\end{APACrefauthors}%
\unskip\
\newblock
\APACrefYearMonthDay{1990}{}{}.
\newblock
\APACrefbtitle {Efficient memory-based learning for robot control} {Efficient
  memory-based learning for robot control}\ \APACbVolEdTR{}{\BTR{}\ \BNUM\
  UCAM-CL-TR-209}.
\newblock
\APACaddressInstitution{}{University of Cambridge, Computer Laboratory}.
\PrintBackRefs{\CurrentBib}

\bibitem [\protect \citeauthoryear {%
Mora%
\ \BBA {} Bialek%
}{%
Mora%
\ \BBA {} Bialek%
}{%
{\protect \APACyear {2011}}%
}]{%
mora_are_2011}
\APACinsertmetastar {%
mora_are_2011}%
\begin{APACrefauthors}%
Mora, T.%
\BCBT {}\ \BBA {} Bialek, W.%
\end{APACrefauthors}%
\unskip\
\newblock
\APACrefYearMonthDay{2011}{}{}.
\newblock
{\BBOQ}\APACrefatitle {Are biological systems poised at criticality?} {Are
  biological systems poised at criticality?}{\BBCQ}
\newblock
\APACjournalVolNumPages{Journal of Statistical Physics}{144}{2}{268--302}.
\PrintBackRefs{\CurrentBib}

\bibitem [\protect \citeauthoryear {%
Timme%
, Alford%
, Flecker%
\BCBL {}\ \BBA {} Beggs%
}{%
Timme%
\ \protect \BOthers {.}}{%
{\protect \APACyear {2014}}%
}]{%
timme_synergy_2014}
\APACinsertmetastar {%
timme_synergy_2014}%
\begin{APACrefauthors}%
Timme, N.%
, Alford, W.%
, Flecker, B.%
\BCBL {}\ \BBA {} Beggs, J\BPBI M.%
\end{APACrefauthors}%
\unskip\
\newblock
\APACrefYearMonthDay{2014}{{\APACmonth{04}}}{}.
\newblock
{\BBOQ}\APACrefatitle {Synergy, redundancy, and multivariate information
  measures: an experimentalist's perspective} {Synergy, redundancy, and
  multivariate information measures: an experimentalist's perspective}.{\BBCQ}
\newblock
\APACjournalVolNumPages{Journal of Computational
  Neuroscience}{36}{2}{119--140}.
\newblock
\begin{APACrefDOI} \doi{10.1007/s10827-013-0458-4} \end{APACrefDOI}
\PrintBackRefs{\CurrentBib}

\bibitem [\protect \citeauthoryear {%
Van~Orden%
, Hollis%
\BCBL {}\ \BBA {} Wallot%
}{%
Van~Orden%
\ \protect \BOthers {.}}{%
{\protect \APACyear {2012}}%
}]{%
van_orden_blue-collar_2012}
\APACinsertmetastar {%
van_orden_blue-collar_2012}%
\begin{APACrefauthors}%
Van~Orden, G.%
, Hollis, G.%
\BCBL {}\ \BBA {} Wallot, S.%
\end{APACrefauthors}%
\unskip\
\newblock
\APACrefYearMonthDay{2012}{}{}.
\newblock
{\BBOQ}\APACrefatitle {The blue-collar brain} {The blue-collar brain}.{\BBCQ}
\newblock
\APACjournalVolNumPages{Fractal Physiology}{3}{}{207}.
\newblock
\begin{APACrefDOI} \doi{10.3389/fphys.2012.00207} \end{APACrefDOI}
\PrintBackRefs{\CurrentBib}

\bibitem [\protect \citeauthoryear {%
Wagenmakers%
, van~der Maas%
\BCBL {}\ \BBA {} Farrell%
}{%
Wagenmakers%
\ \protect \BOthers {.}}{%
{\protect \APACyear {2012}}%
}]{%
wagenmakers_abstract_2012}
\APACinsertmetastar {%
wagenmakers_abstract_2012}%
\begin{APACrefauthors}%
Wagenmakers, E\BPBI J.%
, van~der Maas, H\BPBI L\BPBI J.%
\BCBL {}\ \BBA {} Farrell, S.%
\end{APACrefauthors}%
\unskip\
\newblock
\APACrefYearMonthDay{2012}{}{}.
\newblock
{\BBOQ}\APACrefatitle {Abstract {Concepts} {Require} {Concrete} {Models}}
  {Abstract {Concepts} {Require} {Concrete} {Models}}.{\BBCQ}
\newblock
\APACjournalVolNumPages{Topics in Cognitive Science}{4}{1}{87--93}.
\newblock
\begin{APACrefDOI} \doi{10.1111/j.1756-8765.2011.01164.x} \end{APACrefDOI}
\PrintBackRefs{\CurrentBib}

\bibitem [\protect \citeauthoryear {%
Williams%
\ \BBA {} Beer%
}{%
Williams%
\ \BBA {} Beer%
}{%
{\protect \APACyear {2011}}%
}]{%
williams_generalized_2011}
\APACinsertmetastar {%
williams_generalized_2011}%
\begin{APACrefauthors}%
Williams, P\BPBI L.%
\BCBT {}\ \BBA {} Beer, R\BPBI D.%
\end{APACrefauthors}%
\unskip\
\newblock
\APACrefYearMonthDay{2011}{}{}.
\newblock
{\BBOQ}\APACrefatitle {Generalized {Measures} of {Information} {Transfer}}
  {Generalized {Measures} of {Information} {Transfer}}.{\BBCQ}
\newblock
\APACjournalVolNumPages{arXiv:1102.1507 [physics]}{}{}{}.
\newblock
\APACrefnote{arXiv: 1102.1507}
\PrintBackRefs{\CurrentBib}

\end{thebibliography}

\end{document}